\documentclass[twocolumn,aps,prl,superscriptaddress,notitlepage,longbibliography,reprint]{revtex4-2}

\usepackage{graphicx}
\usepackage{epsfig}
\usepackage{float}
\usepackage{dcolumn}
\usepackage{bm}
\usepackage{amsmath}
\usepackage{float}
\usepackage{amssymb}
\usepackage{lipsum}
\usepackage{titletoc}
\usepackage{makecell}
\usepackage[colorlinks,linkcolor=blue]{hyperref}
\usepackage{array}
\usepackage{multirow}
\usepackage{color}
\usepackage{bbding}
\usepackage{stackengine}

\newcommand{\be}{\begin{equation}}
\newcommand{\ee}{\end{equation}}

\newcommand{\bea}{\begin{eqnarray}}
\newcommand{\eea}{\end{eqnarray}}

\begin{document}

\title{Induced Zeeman effect of moir\'e surface states
in topological insulators}
\author{Haijiao Ji} 
\author{Noah F. Q. Yuan}
\email{fyuanaa@connect.ust.hk}
\affiliation{Tsung-Dao Lee Institute, Shanghai Jiao Tong University, Shanghai 201210, China}
\affiliation{School of Physics and Astronomy, Shanghai Jiao Tong University, Shanghai 200240, China}
\affiliation{Shanghai Research Center for Quantum Sciences, Shanghai 201315, China}

\begin{abstract}
Recently, moiré superlattices have been found on the surface of topological insulators due to the rotational misalignment of topmost layers. 
In this work, we study the effects of moiré superlattices on the Landau levels of topological surface states. 
We find that an extra Zeeman term besides the intrinsic one can be induced by the orbital effect of the magnetic field in moir\'e surface states. 
As a result, the originally field-independent zeroth Landau level of moir\'e surface states could be tilted by the out-of-plane magnetic field.
\end{abstract}
\maketitle

\textit{\textcolor{blue}{Introduction.}}---
It is said that an external magnetic field has two effects on an electron, the orbital effect that couples to momentum, and the Zeeman effect that couples to spin. The orbital effect can be inherited from classical electrodynamics, while the origin of the Zeeman effect was unknown until 1928, when Paul Dirac conceived his famous relativistic equation for electrons \cite{D1,D2}. 

It is derived from the Dirac equation that, microscopically the Zeeman effect is induced by the orbital effect through coupling between electrons and holes \cite{D1,D2}. For clearness, in the rest of this manuscript, we refer to the Zeeman effect that derived from the fundamental Dirac equation as the \textit{intrinsic} Zeeman effect.

In recent developments of condensed matter physics, one important trend is to mimic fundamental principles in condensed matter systems (such as the Dirac equation in graphene \cite{graphene1,graphene2} and topological insulators \cite{TI0,TI1,TI2,TI3}), 
where we physicists are able to tune related parameters \cite{tune1,tune2,tune3,tune4,tune5}, and search for exotic physics, in particular in topological materials \cite{TI1,TI2,TI3,TI4,exotic1,exotic2}.

\begin{figure}
\includegraphics[width=1\columnwidth]{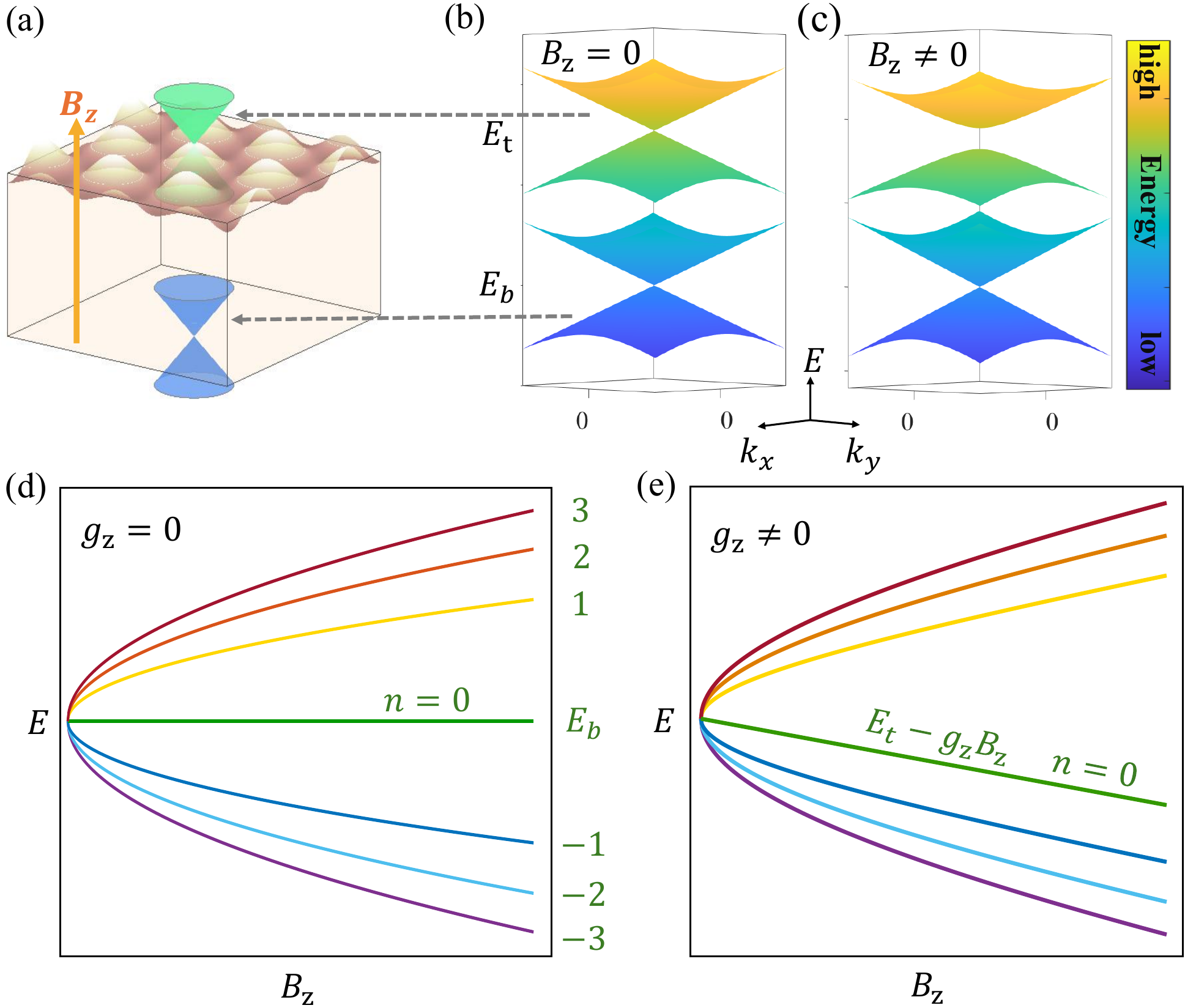}
\caption{ (a) A schematic illustration of the moiré superlattice on the top surface of a topological insulator. The green/blue Dirac cones represent the top and bottom surface states, and the orange arrow denotes an out-of-plane external magnetic field $B_z$.
(b,c) The band structure of the moiré surface states without (b) and with (c) magnetic field $B_z$. The moiré unit cell size is 2$\times$2 of the original lattice, the moiré potential is Eq. (\ref{eq_comb}) with $U_0=0.8$, the bulk contains 5 layers, and the magnetic flux of moiré unit cell is $\frac{1}{3}$ of the flux quantum.
(d,e) Surface state Landau levels with (d) $g_z=0$ and (e) $g_z\neq 0$.
}\label{fig1}
\end{figure}

The topological surface states of a topological insulator can be described by the massless Dirac equation in two dimensions. 
Recently, moiré superlattices have been found on the surface of topological insulators due to the rotational misalignment of topmost layers, which can be modeled by the massless Dirac fermion in a moiré superlattice potential as shown in Fig. \ref{fig1}(a).
The so-called moiré surface states will be formed, as proposed and studied by T. Wang \textit{et. al} \cite{MSS1} and J. Cano \textit{et. al} \cite{MSS2} independently, where multiple Dirac points are formed due to band folding and topological protection.

In this work, we are going to show that, in moiré surface states, through the mechanism of Dirac point coupling, an extra Zeeman term besides the intrinsic one can be induced by the orbital effect of the magnetic field.
We first derive the induced Zeeman effect at the $\Gamma$ point analytically, and work out the corresponding Landau levels.
We then carry out numerical simulations of moiré surface states, where the intrinsic Zeeman effect is neglected.
For comparison, we assume the moiré superlattice is formed on the top layer but not the bottom layer of the topological insulator, as shown in Fig. \ref{fig1}(a), and corresponding Dirac cones of two surfaces are as shown in Fig. \ref{fig1}(b). 
Under an out-of-plane magnetic field, top-layer surface states affected by the moiré potential will be gapped out, while the bottom-layer surface states without moiré superlattice will not, as shown in Fig. \ref{fig1}(c).
Consequently, the Landau levels without and with the induced Zeeman effect are plotted in Figs. \ref{fig1}(d) and (e), respectively.
In the end, we discuss possible realizations and material candidates in experiments.




\textit{\textcolor{blue}{Moiré surface states under magnetic field}}---
We start with the moiré surface states at zero field, which can be regarded as a massless Dirac fermion in a scalar potential 
\be\label{eq_H0}
\mathcal{H}=\hbar v_{\rm F}(\bm\sigma\times\bm k)\cdot\hat{\bm z} + U(\bm r)\sigma_0,
\ee
where $v_{\rm F}$ is the Fermi velocity, $\bm k=(k_x,k_y)$ is the two-dimensional (2D) momentum, $\bm\sigma=(\sigma_x,\sigma_y,\sigma_z)$ are the Pauli matrices denoting spin, and $\sigma_0$ is the 2 by 2 identity matrix.
Now we allow the continuous translation symmetry to be broken into discrete symmetries by the moiré superlattice, while we leave the time-reversal symmetry intact. Then the lowest-order perturbation can be described by a spin-independent periodic scalar potential $U(\bm r+\bm L_{1,2})=U(\bm r)$ with $\bm L_{1,2}$ as the two primitive vectors of the moiré superlattice. A schematic diagram of this setup is shown in Fig. \ref{fig1}(a). This model can apply to bulk TI crystals with top layers twisted or the surface state in the interface between a topological insulator and a large-gap insulator.

\begin{figure}
\includegraphics[width=1\columnwidth]{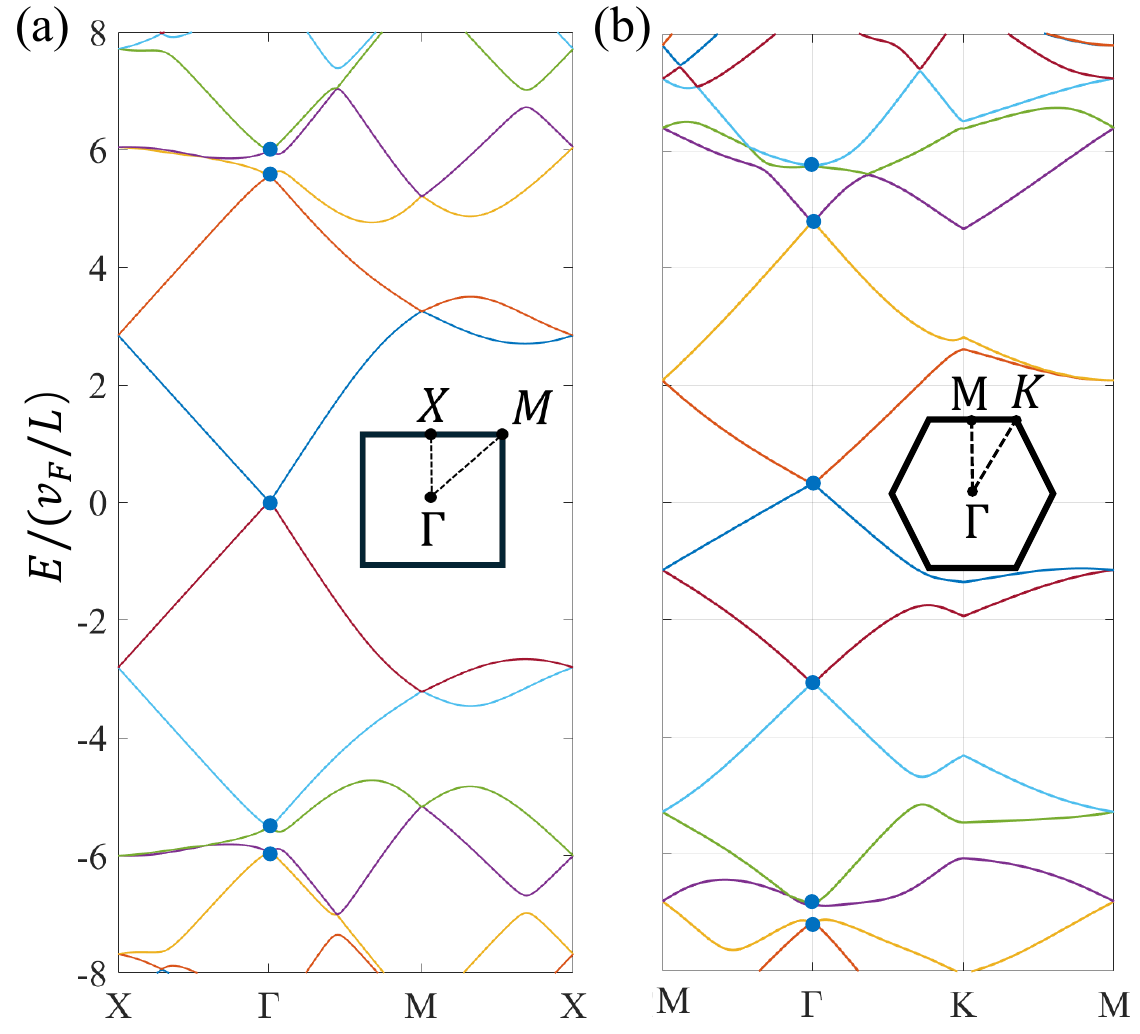}
\caption{The spectrum and multiple Dirac points of (a) square and (b) triangular lattice under moiré potential Eq. (\ref{eq_UG}) with $U = v_{F}/L$. The corresponding mini Brillouin zone is shown in the center. The entire spectrum remains gapless due to the symmetry anomaly. The energy spectrum of the square lattice maintains electron-hole symmetry, while that of the triangular lattice breaks it. In both cases, multiple Dirac points are formed at the $\Gamma$ point (highlighted in blue dots).}\label{fig2}
\end{figure}

The topological surface states could not be gapped out by a time-reversal invariant scalar potential due to topological protection \cite{MSS1,MSS2}.  
Multiple Dirac points of the moiré surface states are shown in Fig. \ref{fig2}, where the moir\'e potential are chosen as leading-order Fourier series \cite{MSS1}
\be\label{eq_UG}
U(\bm r)=2U\sum_{i}\cos(\bm G_i\cdot\bm r),
\ee
where $\bm G_i$ are reciprocal vectors of the moir\'e superlattice related by the underlying rotation symmetry.
The energy spectra of moir\'e surface states are as shown in Fig. \ref{fig2}. 
For a square lattice in (a), in Eq. (\ref{eq_UG}) we have two reciprocal vectors $\bm G_1=2\pi/L(1,0)$ and $\bm G_2=2\pi/L(0,1)$, the moir\'e surface state is particle-hole symmetric due to the emergent symmetry $U(x+L/2,y+L/2)=-U(x,y)$. 
On the other hand, for a triangular lattice in (b), in Eq. (\ref{eq_UG}) we have three reciprocal vectors $\bm G_1=2\pi/L(0,1)$ and $\bm G_{2,3}=2\pi/L(\pm\frac{\sqrt{3}}{2},-\frac{1}{2})$, the moir\'e surface state is particle-hole asymmetric as there is no such emergent symmetry between positive and negative potentials.

At the $\Gamma$ point, the original Dirac point will be present with renormalized velocity \cite{MSS2}, which is at energy $E_0$. As discussed previously, $E_0=0$ for a square lattice in Fig. \ref{fig2}(a) and $E_0\neq 0$ for a triangular lattice in Fig. \ref{fig2}(b). Additional Dirac points will also arise at $\Gamma$ point, which are highlighted in blue dots.

Next, we apply an external magnetic field to the moiré surface states, and focus on the Dirac points at the $\Gamma$ point.
For this purpose, we rewrite the full Hamiltonian of moiré surface states near $\Gamma$ point in the basis of $\Gamma$ point eigenstate doublets (i.e. Dirac doublets)
\be\label{eq_H}
\mathcal{H}_{\Gamma}=
\begin{pmatrix}
    H & V\\
    V^{\dagger} & H'
\end{pmatrix},
\ee
where we are interested in the Dirac point at $E_0$ as described by the 2 by 2 Dirac Hamiltonian $H$, the rest Dirac points are included in Hamiltonian $H'$, and the coupling between $H$ and $H'$ is denoted as $V$.
At $k=0$, by definition Eq. (\ref{eq_H}) should be fully diagonal to a set of doublets $\mathcal{H}_{\Gamma}^{(0)}={\rm diag}(E_0,E_1,E_2,\dots,E_l,\dots)\otimes\sigma_0$, where $\{E_l\}$ are Dirac point eigenenergies.
For concreteness, with each doublet of energy $E_l$, we choose the basis such that the time-reversal symmetry is represented by $\mathcal{T}=i\sigma_y\mathcal{K}$, with complex conjugate operator $\mathcal{K}$.
Thus, 
up to the linear order in $\bm k$, we have the expansions 
\bea\label{eq_D}
H&=&E_0+\hbar v_0(\bm\sigma\times\bm k)\cdot\hat{\bm z} + O(k^2),\\
V&=&(V_1,V_2,\dots,V_l,\dots),\\ \label{eq_Vl}
V_l&=&V_l^{\alpha\beta}\sigma_{\alpha}k_{\beta}+O(k^2),
\eea
where the renormalized Fermi velocity is $v_0$ \cite{MSS2}, and the Dirac point couplings are described by complex coefficients $\{V_{l}^{\alpha\beta}\}$. 
It is important to note that, in Eq. (\ref{eq_Vl}), the Einstein summation is assumed, and the Pauli matrices $\sigma_{\alpha}$ and momentum $k_{\beta}$ are coupled since $\sigma_{\alpha},k_{\beta}$ are both odd while the Hamiltonian $V$ is even under the time-reversal symmetry $\mathcal{T}=i\sigma_y\mathcal{K}$. 

The general projection method leads to the effective Hamiltonian near Dirac point energy $E_0$
\be
H_{\rm eff}=H+V(E_0-H')^{-1}V^{\dagger}.
\ee
To the leading order, $V(E_0-H')^{-1}V^{\dagger}$ is of the order of $k^2$ at zero field. 
Under an external magnetic field $\bm B$, we make 
the minimal coupling
\be
\bm k\to\bm\pi\equiv\bm k-\frac{e}{\hbar}\bm A,
\ee
then the correction $V(E_0-H')^{-1}V^{\dagger}$ will be a quadratic form of the canonical momentum operator $\bm\pi$. 
Here $e$ is the electron charge, and $\bm A$ is the vector potential. 
Notice that we do not introduce intrinsic Zeeman effect here.

Unlike momentum $\bm k$ (pure numbers), the canonical momentum components do not commute, whose commutator is proportional to the magnetic field strength
\be\label{eq_cm}
[\pi_{\mu},\pi_{\nu}]=i\frac{e}{\hbar}\epsilon_{\mu\nu\lambda}B_{\lambda},
\ee
where $\epsilon_{\mu\nu\lambda}$ is the totally antisymmetric rank-3 tensor.

The commutator Eq. (\ref{eq_cm}) converses the second order term of $\bm\pi$ on the left hand side to the zeroth order term on the right hand side. Thus, up to the linear order in $\bm\pi$, we find the Zeeman effect correction, which is of the zeroth order of $\bm\pi$, in the effective Hamiltonian
\be\label{eq_eff}
H_{\rm eff}=E_0+\hbar v_0(\bm\sigma\times\bm\pi)\cdot\hat{\bm z}+g_{\mu\nu}\sigma_{\mu}B_{\nu},
\ee
where the induced $g$-factor $g_{\mu\nu}$ is related to the Dirac coupling coefficients $\{V_{l}^{\alpha\beta}\}$ as follows
\be\label{eq_g}
g_{\mu\nu}=\frac{e}{\hbar}\sum_{l\neq 0}\sum_{\alpha\alpha'\beta\beta'}
\frac{\epsilon_{\alpha\alpha'\mu}\epsilon_{\beta\beta'\nu}}{E_l-E_0}{\rm Re}\left(V_{l}^{\alpha\beta}\overline{V}_{l}^{\alpha'\beta'}\right).
\ee
Here $\overline{z}$ denotes the complex conjugate of complex number $z$, and Einstein summation is assumed in Eq. (\ref{eq_eff}).

We have worked out the induced Zeeman term of the Dirac point at energy $E_0$ in Eqs. (\ref{eq_eff},\ref{eq_g}). 
In the $g$-factor expression of Eq. (\ref{eq_g}), the orbital effect is reflected by the flux quantum $\phi_0=h/e$, and the Dirac point coupling is described by $\{V_{l}^{\alpha\beta}\}$.
The same mechanism and formula can also be applied to other Dirac points at $\Gamma$ point or other time-reversal invariant points, as long as multiple Dirac points coexist.

To make the induced $g$-factor nonzero, as shown in Eq. (\ref{eq_g}), one needs to break the particle-hole symmetry of $\mathcal{H}$ with respect to $E_0$. 
This can be realized in a triangular lattice as shown Fig. \ref{fig2}(b) whose energy spectrum is particle-hole asymmetric.

As shown in Fig. \ref{fig2}(a) for a square lattice, moir\'e surface state under the moir\'e potential Eq. (\ref{eq_UG}) composed of first-order harmonics is particle-hole symmetric, and hence the induced $g$-factor is zero in this case. 
To break the emergent particle-hole symmetry, one may include higher order harmonics in moir\'e potential besides Eq. (\ref{eq_UG}). Alternatively, one can also employ the Dirac comb potential for $U(\bm r)$ as elaborated in the session of tight-binding simulation.

Our model in Eq. (\ref{eq_H0}) contains a massless Dirac fermion which has particle-hole symmetry, and the particle-hole asymmetry has to be from moir\'e potential.
However, our derivation of induced $g$-factor is based on $\Gamma$-point Hamiltonian in Eq. (\ref{eq_H}) and does not rely on a particle-hole symmetric Dirac fermion.
In realistic materials, the surface state is beyond the simple Dirac model, and the particle-hole symmetry could be broken without further constraints for the moir\'e potential, where moir\'e-induced Zeeman effect is widely expected. 

\textit{\textcolor{blue}{Symmetry analysis and Landau levels}}---
Next, we carry out the symmetry analysis of the Zeeman term $H_{Z}\equiv g_{\mu\nu}\sigma_{\mu}B_{\nu}$ and hence work out corresponding Landau levels.
The Dirac Hamiltonian Eq. (\ref{eq_D}) has the point group $C_{nv}$ with $n=3,4,6,\infty$.
Under point group $C_{nv}$, the invariant Zeeman term takes the following form \cite{TI0}
\be
H_Z=g_{\parallel}(\sigma_xB_x+\sigma_yB_y)+g_z\sigma_z B_z
\ee
with two $g$-factors due to Dirac point coupling
\bea
g_{\parallel}=\frac{e}{\hbar}\sum_{l}
\frac{2}{E_l-E_0}{\rm Re}\left(V_{l}^{zx}\overline{V}_{l}^{xy}-V_{l}^{zy}\overline{V}_{l}^{xx}\right),\\
g_{z}=\frac{e}{\hbar}\sum_{l}
\frac{2}{E_l-E_0}{\rm Re}\left(V_{l}^{xx}\overline{V}_{l}^{yy}-V_{l}^{xy}\overline{V}_{l}^{yx}\right).
\eea
The Zeeman term $H_Z$ could contain diagonal $(g_zB_z)$ or off-diagonal $(g_{\parallel}B_{x,y})$ terms. We can absorb the off-diagonal terms $\bm h\equiv g_{\parallel}(B_x,B_y)$ into the redefinition of $\bm\pi$, namely $\hbar v_0\tilde{\bm\pi}=\hbar v_0\bm\pi+\bm h\times\hat{\bm z}$. Such redefinition does not alter the commutator Eq. (\ref{eq_cm}) since $h_{x,y}$ are pure numbers and commute with $\pi_{x,y}$. For simplicity, we thus ignore the tilde symbol in $\tilde{\bm\pi}$ for the following calculations.

Then we only have the diagonal term $g_zB_z$ to deal with, and the effective Hamiltonian Eq. (\ref{eq_eff}) becomes
\be
H_{\rm eff}=E_0+
\begin{pmatrix}
    g_zB_z & \mathcal{E}a \\
    \mathcal{E}a^{\dagger} & -g_zB_z
\end{pmatrix}
\ee
where $\mathcal{E}=v_0\sqrt{2eB_z\hbar}$ and the annihilation operator is
\be
a=\frac{\pi_y-i\pi_x}{\sqrt{2eB_z/\hbar}},\quad
[a,a^{\dagger}]=1.
\ee
We seek the eigenstate of the following ansatz
\be
\psi_n=
\begin{pmatrix}
    u|n-1\rangle\\
    v|n\rangle
\end{pmatrix},
\ee
where $u,v$ are complex parameters to be determined later, and $|n\rangle$ are given by
$a|0\rangle=0,$ $|n\rangle=\frac{1}{\sqrt{n}}a^{\dagger}|n-1\rangle$
with $n=1,2,\dots$. Notice that $|n\rangle=0$ when $n<0$.
The eigenstate equation $H_{\rm eff}\psi_n=E\psi_n$ can be reduced to
\be
\begin{pmatrix}
    g_zB_z & \mathcal{E}\sqrt{n}\\
    \mathcal{E}\sqrt{n} & -g_zB_z
\end{pmatrix}
\begin{pmatrix}
    u\\
    v
\end{pmatrix}
=(E-E_0)
\begin{pmatrix}
    u\\
    v
\end{pmatrix},
\ee
and the Landau levels can then be worked out.

As a result, the Landau levels of Dirac point at $E_0$ are only affected by the out-of-plane magnetic field $B_z$ but not the in-plane components $B_{x,y}$
\bea
n\neq 0&:&\ E=E_0+{\rm sgn}(n)\sqrt{2ev_0^2|n|B_z\hbar+g_z^2B_z^2},\\
n=0&:&\ E=E_0-g_zB_z.
\eea
Landau levels of the Dirac point with $g_z=0$ and $g_z\neq 0$ are plotted in Figs. \ref{fig1}(d) and (e) respectively. 
With $g_z=0$, Landau levels are symmetric with respect to the field-independent zeroth Landau level $E=E_0$ [Fig. \ref{fig1}(d)], while with $g_z\neq 0$, Landau levels become asymmetric and the zeroth Landau level is tilted by the field $E=E_0-g_zB_z$ [Fig. \ref{fig1}(e)].

To close this session, we classify the Dirac points according to the point group $C_{nv}$.
Including spin, the point group becomes the double group $C_{nv}^{*}$. We can find the two-dimensional (2D) irreducible representations of $C_{nv}^{*}$ with basis states labeled by total angular momenta $\pm j$ with the possible values of $j$ as $(n=3,4,6)$
\be
j= \pm\frac{1}{2},\quad\pm 1,\quad\dots,\quad\pm\frac{n-1}{2}.
\ee
Notice that under $n$-fold rotation, $j$ is equivalent to $j\pm n$, hence $j=\pm n/2$ are equivalent and do not form a 2D irreducible representation. As a result, $j\leq (n-1)/2$.

Dirac points are 2D irreducible representations of $C_{nv}^{*}$ and are therefore labeled by total angular momentum $j$. The Dirac point energies are labeled by $E_l$, where $l$ can be related to the orbital angular momentum \cite{MSS2}. 


\textit{\textcolor{blue}{Tight-binding model simulation}}---
In this section, we consider the numerical simulation of the analytical calculations for the induced Zeeman effect in previous sessions. 
We have used the $k\cdot p$ model in Eq. (\ref{eq_H0}) to calculate the energy spectrum of moir\'e surface states at zero field. Under a uniform magnetic field, the vector potential $\bm A$ could become divergent at long distances and hence the $k\cdot p$ model in Eq. (\ref{eq_H0}) may not converge properly. As a result, we choose the tight-binding model to carry out our numerical simulations.

Notice that topological surface states cannot be simulated by 2D tight-binding models due to topological obstructions \cite{s1,s2,s3,s4,s5}. 
For this reason, we consider the moiré surface states on the top surface of a three-dimensional (3D) topological insulator, realized in a 3D lattice
\begin{equation}
H(\boldsymbol{k},\bm r)=H_{\rm BHZ}(\boldsymbol{k})+U(\boldsymbol{r})\delta(z),
\end{equation}
where $U(\bm r+\bm L_{1,2})=U(\bm r)$ denotes the periodic scalar potential in real space $\bm r=(x,y,z)$ and $\bm L_{1,2}$ are two primitive vectors of the moiré superlattice on the top surface $z=0$. 
The 2D moiré potential can be modeled by the 2D Dirac comb function
\be\label{eq_comb}
U(\bm r)=U_0|\bm L_1\times\bm L_2|\sum_{n_1,n_2\in\mathbb{Z}}\delta^2(\bm r-n_1\bm L_1-n_2\bm L_2),
\ee
where particle-hole symmetry is broken and the induced $g$-factor is expected nonzero.
The Hamiltonian $H_{\rm BHZ}(\bm k)$ is the Bernevig-Hughes-Zhang (BHZ) model realized in a 3D cubic lattice with momentum $\bm k=(k_x,k_y,k_z)$,
\begin{equation}
H_{\rm BHZ}=M\tau_z+t\{\sin(k_z)\sigma_z+\sin(k_x)\sigma_y-\sin(k_y)\sigma_x\}\tau_x,
\label{eq_BHZ4}
\end{equation} 
in the basis $\left( c^\dagger_{1\uparrow,\mathbf{k}},\; c^\dagger_{1\downarrow,\mathbf{k}},\; c^\dagger_{2\uparrow,\mathbf{k}},\; c^\dagger_{2\downarrow,\mathbf{k}} \right)$, where the Pauli matrices $\sigma_i$, $\tau_i$ ($i = {x, y, z}$) act in the spin ($\uparrow$ or $\downarrow$) and orbital (1 or 2) spaces respectively.  
The Dirac mass term is $M=m_0+3m_1-m_1(\cos k_x+\cos k_y+\cos k_z)$, with parameter $m_0/m_1$ controlling the topological phase transition.
The system is topological when $-6<m_0/m_1<0$.
Bulk Hamiltonian Eq. (\ref{eq_BHZ4}) respects the point group $D_{4h}$, and the corresponding surface Hamiltonian respects the point group $C_{4v}$.
As illustrated in Fig. \ref{fig1}, we work out the band structure of the 3D BHZ model with the moiré potential on the top layer. 
At zero field, to the leading order, the moiré potential changes the Dirac point energy of the top layer, as shown in Fig. \ref{fig1}(b). We denote the Dirac point energies of the top and bottom surfaces as $E_t$ and $E_b$, respectively.
Then an external magnetic field $\mathbf{B}$ is applied by the Peierls substitution \cite{TBA1,TBA2,TBA3,TBA4}, 
\begin{equation}
    t_{ij} \rightarrow t_{ij} \exp\left[ i \frac{e}{\hbar} \int_{\mathbf{r}_i}^{\mathbf{r}_j} d\mathbf{r} \cdot \mathbf{A}(\mathbf{r}) \right],
    \label{eq:peierls}
\end{equation}
where $\mathbf{r}_{i,j}$ denote the lattice sites and $t_{ij}$ denotes the hopping integral from $\mathbf{r}_{i}$ to $\mathbf{r}_{j}$. 
In our numerical simulation, we assume the intrinsic Zeeman effect is zero.
Fig. \ref{fig1}(c) depicts the two Dirac cones under $\mathbf{B}=B_z\hat{\bm z}$. 
The generic theoretical analysis of induced Zeeman effect can be applied to moiré superlattices of topological materials, such as Bi$_2$Se$_3$, Bi$_2$Te$_3$ and Sb$_2$Te$_3$ family \cite{Miao2025,He2020,BiTe1,BiTe2,BiTe3,LCX2024}, and transition metal dichalcogenide family \cite{mote1,mote2,mote3,mote4,mote5}. 
One may use the tight-binding model
$H_{\rm BHZ}=M\tau_z+\{t_z\sin(k_z)\sigma_z+S_{+}\sigma_{+}+S_{-}\sigma_{-}\}\tau_x$ to simulate triangular lattices,
where the Dirac mass term becomes $M=m_0+3m_2-m_2[C(\boldsymbol{k})+\cos({k_z})]$, with $C(\boldsymbol{k})=\sum_{j=1}^{3} \cos \left(\boldsymbol{k} \cdot \boldsymbol{R}_{j}\right)$ and $S_{+}(\bm k)=S_{-}^{*}(\bm k)=\sum_{j=1}^{3} \omega^{j-1} \sin \left(\bm k \cdot \boldsymbol{R}_{j}\right)$. 
The bonding vectors of the nearest sites are $\boldsymbol{R}_{1}=L\boldsymbol{x}$, $\boldsymbol{R}_{2}=   L(-\boldsymbol{x} / 2+\sqrt{3} \boldsymbol{y} / 2)$, $\boldsymbol{R}_{3}=L(-\boldsymbol{x} / 2-\sqrt{3} \boldsymbol{y} / 2)$  with lattice constant $L$, and $\omega=\exp(2\pi i/3)$ is the cubic root of unity. 
The system is topological when $-5<m_0/m_2<-3$ and $-1<m_0/m_2<1$.
The bulk then respects point group $D_{3d}$, and the surface $C_{3v}$.


\textit{\textcolor{blue}{Conclusion}}---
In this work, we show the generic mechanism of induced Zeeman effect in moiré surface states, as the combined consequence of orbital effect and coupling between multiple Dirac points. Our theory treats the magnetic field as a perturbation to the moir\'e potential. When the magnetic field is large, the superlattice potential is instead a perturbation, and we may need to consider the broadening of Landau levels into Chern bands with finite bandwidth \cite{Pfannkuche1992,moireLL1,moireLL2,moireLL3}.

\textit{\textcolor{blue}{Acknowledgements.}}---
We thank Hao Zheng and Yi Zhang for fruitful discussions.
The authors thank Chao-Xing Liu for inspiring discussions.
This work is supported by the National Natural Science Foundation of China (Grant. No. 12174021) and Cultivation Project of Shanghai Research Center for Quantum Sciences (Grant No. LZPY2024).

\end{document}